\documentclass[conference]{IEEEtran}
\usepackage{graphicx}

\ifCLASSINFOpdf
\else
\fi
\begin{document}
%
\title{Integrating ``self-efficacy" into a gamified approach to thwart phishing attacks}

\author{\IEEEauthorblockN{Nalin Asanka Gamagedara Arachchilage and Mumtaz Abdul Hameed}
\IEEEauthorblockA{Australian Centre for Cyber Security\\
The School of Engineering and Information Technology\\
University of New South Wales\\
Australian Defence Force Academy\\
PO Box 7916 Canberra BC ACT 2610\\
Email: nalin.asanka@adfa.edu.au}
}


%


\maketitle

\begin{abstract}
 Security exploits can include cyber threats such as computer programs that can disturb the normal behavior of computer systems (viruses), unsolicited e-mail (spam), malicious software (malware), monitoring software (spyware), attempting to make computer resources unavailable to their intended users (Distributed Denial-of-Service or DDoS attack), the social engineering, and online identity theft (phishing). One such cyber threat, which is particularly dangerous to computer users is phishing. Phishing is well known as online identity theft, which targets to steal victims’ sensitive information such as username, password and online banking details. Automated anti-phishing web browser plugin tools have been developed and used to alert users of potential fake emails and websites. However, these tools are not completely reliable in detecting and protecting people from phishing attacks. This is because the ``humans are the weakest link" in information security. It is not possible to completely circumvent the end-user, for example, in personal computer use, one mitigating approach for computer and information security is to educate the end-user in security prevention. Educational researchers and industry experts talk about well-designed user security education can be effective. However, we know to our cost no-one talks about how to better design security education (i.e. user-centered security education) for end-users. Therefore, this paper focuses on designing an innovative and gamified approach to educate individuals about phishing attacks. The study asks how one can integrate ``self-efficacy", which has a co-relation with the user's knowledge, into an anti-phishing educational game to thwart phishing attacks? One of the main reasons would appear to be a lack of user knowledge to prevent from phishing attacks. Therefore, this research investigates the elements that influence (in this case, either conceptual or procedural knowledge or their interaction effect) and then integrate them into an anti-phishing educational game to enhance people's phishing prevention behaviour through their motivation.

\end{abstract}


%
\IEEEpeerreviewmaketitle

\section{Introduction}

In March 2016, John Podesta, the chairman of the Hillary Clinton presidential campaign, received a phishing email with a subject line ``*Someone has your password*'' (shown in Fig. 1). The email greeted Podesta, ``Hi John" and then reads, ``Someone just used your password to try to sign into your Google Account john.podesta@gmail.com''. Then it provided a time stamp (i.e. Saturday, 19 March, 8:34:30 UTC) and an IP address (i.e. 134.249.139.239) in the Location (i.e. Ukraine), where someone used his password. The email also stated ``Google stopped this sign-in attempt". It then offered a link luring him to change his password immediately. Bit.ly \cite{Bitly2017} provides web address shortening service (i.e. this is heavily used by Twitter users), which can make users easier to share. So, hackers created a Bit.ly account called Fancy Bear, belongs to a group of Russian hackers, to by-pass the web browser phishing email filtering system in order to make the attack successful. Unfortunately, Podesta clicked on the link and disclosed his credentials to the hackers. This incident shows us ``humans are the weakest link" in information security. The hacking group, who created the Bit.ly account linked to a domain under the control of Fancy Bear, failed to make the account private. Therefore, this has shown ``hackers are human too and they make mistakes (i.e. weakest link)".

\begin{figure} 
\centering
\includegraphics [height=2.00 in,width=3.25 in]{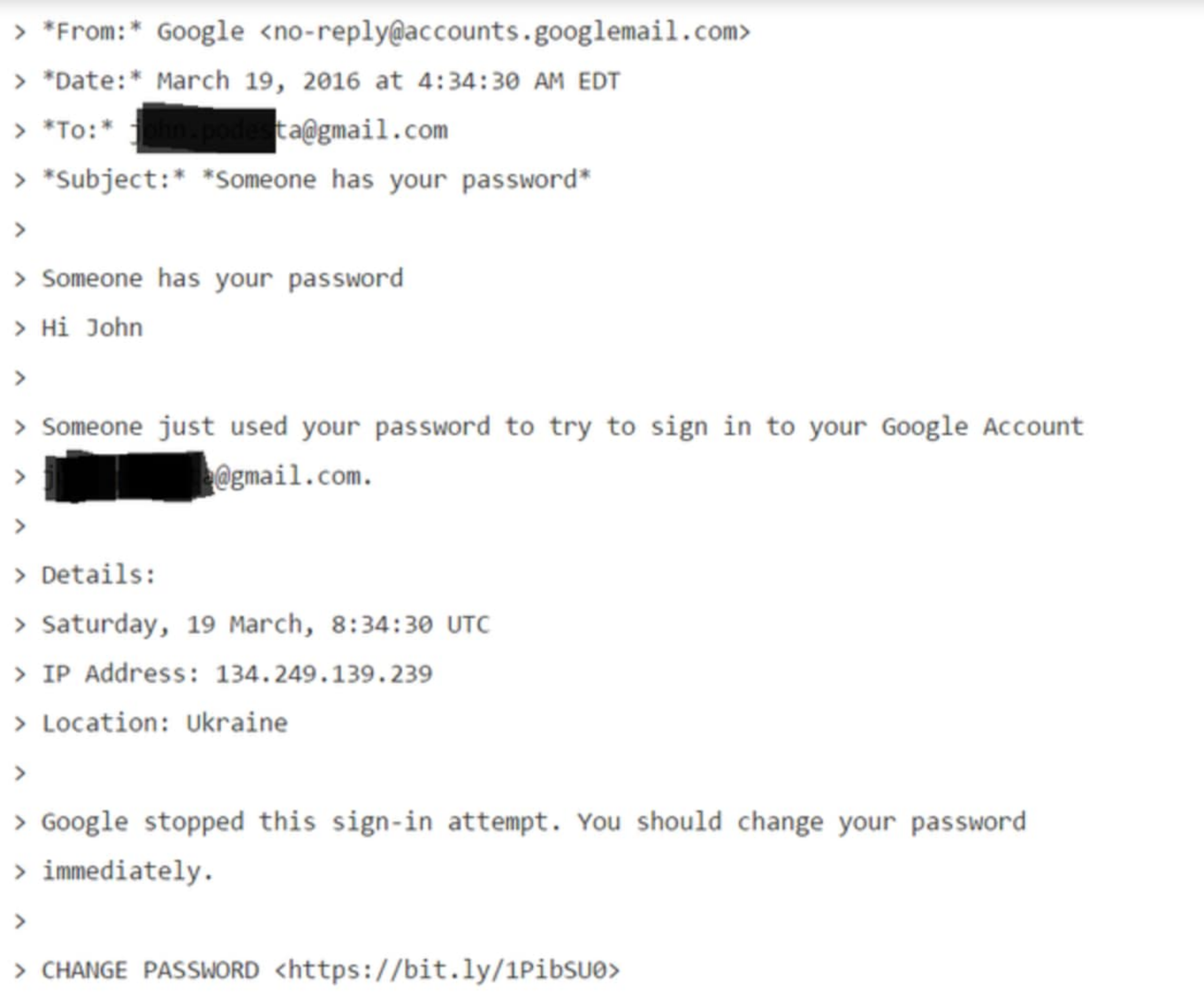}
\caption{The John Podesta emails released by WikiLeaks \cite{Krawchenko2016} \protect}
\vskip -6pt
\end{figure}

Phishing (identity theft) is particularly dangerous to computer users \cite{arachchilage2016phishing} \cite{arachchilage2013game}. It synthesizes social engineering techniques (i.e. the art of human hacking) along with the technical subterfuge to make the attack successful \cite{gupta2017defending}. Phishing aims to steal sensitive information such as username, password and online banking details from its victims \cite{arachchilage2016phishing}. Mass-marketing perpetrators commit identity theft to contact, solicit and obtain money, funds, or other items of value from victims \cite{whitty2015mass}. Online Mass-Marketing Fraud (MMF) is a serious, complex and often very organised crime, which exploits mass communication techniques (e.g. email, Instance messaging service, spams, social networking website) to steal people's money. 

Automated tools (e.g. anti-phishing, anti-virus and anti-spyware) have been developed and used to alert users of potential fraudulent emails and websites. However, these tools are not entirely reliable in detecting and preventing people from online phishing attacks. For example, even the best anti-phishing tools missed over 20 percent of phishing websites \cite{sheng2007anti} \cite{dhamija2006phishing}. Because the ``humans" are the weakest link in cyber security \cite{arachchilage2016phishing}. It is impossible to completely avoid the end-user \cite{arachchilage2013taxonomy}, for example, in personal computer use, one mitigating approach for cyber security is to educate the end-user in security prevention. Educational researchers and industry experts talk about well-designed user security education can be effective \cite{arachchilage2016phishing} \cite{arachchilage2013game} \cite{arachchilage2014security} \cite{sheng2007anti} \cite{kirlappos2012security} \cite{kumaraguru2008lessons}. However, we know to our cost no-one talks about how to better design security education for end-users. Therefore, the aim of this research proposal focuses on designing an innovative and gamified approach to educate individuals about phishing attacks. The study asks how one can integrate ``self-efficacy", which has a co-relation with the user’s knowledge, into an anti-phishing educational game to thwart phishing attacks?

We initially focus on identifying how people's ``self-efficacy" enhances their phishing threat avoidance behaviour. One of the main reasons would appear to be a lack of user knowledge to prevent from phishing attacks. Based on the literature, we then attempt to identify the elements that influence (in this case, either conceptual or procedural knowledge or their interaction effect) to enhance people's phishing prevention behaviour through their motivation. Based on the literature, we will also attempt to identify whether conceptual knowledge or procedural knowledge has a positive effect on computer users' self-efficacy in relation to thwarting phishing attacks. Furthermore, the current research work integrates either procedural or conceptual (or the interaction effect of both procedural and conceptual) knowledge into an anti-phishing educational game to better educate people how to protect themselves against phishing attacks. We believe this will certainly improve users' ability to thwart phishing attacks.

\section{Related Work}

Educational games and simulations have become increasingly acknowledged as an enormous and powerful teaching tool that may result in an ``instructional revolution'' \cite{walls2012using} \cite{cone2007video} \cite{foreman2004video} \cite{arachchilage2012security}. The main reason is that game based education allows users to learn through experience and the use of virtual environment while leading them to approach problem solving through critical thinking \cite{arachchilage2016phishing}. Zyda \cite{zyda2005visual} describes Serious Games as ``a mental contest, played with a computer in accordance with specific rules". Michael and Chen \cite{michael2005serious} proposed the following thematic classification based upon Zyda's definition of Serious Games; Military Games, Government Games, Educational Games, Corporate Games, Healthcare Games, Political and Religious Games. Nevertheless, serious games are quite useful as an effective teaching medium \cite{froschauer2010design}, because it enables users to learn in an interactive and attractive manner \cite{arachchilage2011design}. There is a considerable amount of published studies in the literature describing the role of games in the educational context. Bellotti, et al. \cite{bellotti2009enhancing} designed a Massive Multiplayer Online Game (MMOG) for high-school called `SeaGame' to promote best practices in sea-related behaviours, such as sailing or beach surveillance. The main focus of the `SeaGame' was to embed the educational content into the gaming context in a meaningful, homogeneous and compelling whole, where the player can enjoy learning while having fun. Authors have concluded that this type of games helps people to improve their best practices in behaviour.

In addition, game based education is further useful in motivating players to change their behaviour. Gustafsson and Bang \cite{gustafsson2009evaluation} designed a pervasive mobile-based game called `Power Agent' to educate teenagers and their families to reduce energy consumption in their homes. They attempted to transform the home environment and its devices into a learning arena for hands-on experiences with electricity usage. The results suggested that the game concept was more efficient in engaging teenagers and their families to reduce their energy consumption during the game sessions.
Furthermore, serious games can be effective not only for changing people's behaviour, but also for developing their logical thinking to solve mathematical problems. Eagle and Barnes \cite{eagle2008wu} designed `Wu's Castle' through an empirical investigation, which is a 2D (2-Dimensional) role-playing game where students developed their C++ coding skills such as loops, arrays, algorithms and logical thinking and problem solving. The results showed that `Wu's Castle' is more effective than a traditional programming assignment for learning how to solve problems on arrays and loops.
One of the most well-established facts within the serious games research is the ability to enhance players' motivation towards learning as they are able to retain their attention and keep them engaged and immersed with games \cite{arachchilage2016phishing} \cite{gee2003video} \cite{fotouhi2009mobo} \cite{gunter2008taking} \cite{malone1981toward}. Other interesting facts of educational games are: (a). The provision of immersive gaming environments that can be freely explored by players and promote self-directed learning \cite{oblinger2004next} \cite{squire2003video}. (b). The immediate feedback process with perception of progress \cite{de2006can} \cite{gunter2008taking} \cite{fotouhi2009mobo}. (c). Their relation to constructivist theories and support of learning in the augmented environment \cite{fotouhi2009mobo} \cite{prensky2001digital}.

The design of serious games is a double-edged sword. When its power is properly harnessed to serve good purposes, it has tremendous potential to improve human performance. However, when it is exploited for violation purposes, it can pose huge threats to individuals and society. Therefore, the designing of educational games is not an easy task and there are no all-purpose solutions \cite{moreno2008educational}. The notion that game based education offers the opportunity to embed learning in a natural environment, has repeatedly emerged in the research literature \cite{andrews2003concept} \cite{arachchilage2016phishing} \cite{arachchilage2013game} \cite{arachchilage2014security} \cite{sheng2007anti} \cite{walls2012using} \cite{gee2003video} \cite{roschelle2002walk}. 

A number of educational games have been designed and developed to protect computer users and to assert security issues in the cyberspace \cite{cone2007video}. For example, some educational games teach information assurance concepts, whereas others teach pure entertainment with no basis in information assurance principles or reality. However, there is little research on engagement in the virtual world that also combines the human aspect of security \cite{sheng2007anti}. Therefore, it is worth investigating further on game based learning in order to protect computer users from malicious IT threats such as MMFs.

Even though there are usability experts who claim that user education and training does not work \cite{sheng2007anti}, other researchers have revealed that well-designed end-user education could be a recommended approach to combating against cyber-attacks such as MMFs \cite{allen2006social} \cite{kirlappos2012security} \cite{kumaraguru2008lessons} \cite{schechter2007emperor} \cite{timko2008social}. In line with Herley \cite{herley2009so}, also Kirlappos and Sasse \cite{kirlappos2012security} and other researchers argue that current security education on malicious IT threats offers, little protection to end users, who access potentially malicious websites \cite{kumaraguru2008lessons} \cite{sheng2007anti} \cite{arachchilage2014security}.

Another reason for ineffectiveness of current security education for phishing prevention is because security education providers assume that users are keen to avoid risks and thus likely to adopt behaviours that might protect them. Kirlappos and Sasse \cite{kirlappos2012security} claimed that security education should consider the drivers of end user behaviour rather than warning users of dangers. Therefore, well-designed security education (i.e. user-centred security education) should develop threat perception where users are aware that such a threat is present in the cyberspace. It should also encourage users to enhance avoidance behaviour through motivation to protect them from malicious IT threats.

The literature revealed that well-designed games focusing on education could be helpful for learning even when used without assistance. The `Anti-phishing Phil' game developed by Sheng, et al. \cite{sheng2007anti} reported results confirm that games educate people about phishing and other security attacks in a more effective way than other educational approaches such as reading anti-phishing tutorial or reading existing online training materials. Arachchilage et al. \cite{arachchilage2016phishing} developed a mobile game prototype to teach people how to thwart phishing attacks. Their mobile game design aimed to enhance the user's avoidance behaviour through motivation to protect themselves against phishing threats. The designed mobile game was somewhat effective in teaching people how to thwart phishing attacks as the study results showed a significant improvement of participants' phishing threat avoidance behaviour in their post-test assessment. Furthermore, the study findings suggested that participants' threat perception, safeguard effectiveness, self-efficacy, perceived severity and perceived susceptibility elements positively impact threat avoidance behaviour, whereas safeguard cost had a negative impact on it.

Baslyman and Chiasson \cite{baslyman2016smells} have developed a board game that contributes to enhance users' awareness of online phishing scams. Their findings revealed that after playing the game, participants had a better understanding of phishing scams and learnt how to better protect themselves. Kumaraguru, et al. \cite{kumaraguru2007protecting} designed and evaluated an embedded training email system that teaches people to protect themselves from phishing attacks during their normal use of email. The authors conducted lab experiments contrasting the effectiveness of standard security notices about phishing with two embedded training designs they developed. They found that embedded training works better than the current practice of sending security notices.

Robila and Ragucci \cite{robila2006don} evaluated the impact of end user education in differentiating phishing emails from legitimate ones. They provided an overview of phishing education, targeting on context aware attacks and introduced a new strategy for end user education by combining phishing IQ tests and classroom discussions. The technique involved in displaying both legitimate and fraudulent emails to users and ask them to identify the phishing attempts from the authentic emails. The study concluded that users identified phishing emails correctly after having the phishing IQ test and classroom discussions. Users also acknowledged the usefulness of the IQ test and classroom discussion. Researchers at Indiana University also conducted a similar study of 1700 students in which they collected websites frequently visited by students and either sent them phishing messages or spoofed their email addresses \cite{jagatic2007social}.

Tseng, et al. \cite{tseng2011automatic} also developed a game to teach users about phishing based on the content of the website. The authors proposed the phishing attack frame hierarchy to describe stereotype features of phishing attack techniques. The inheritance and instantiation properties of the frame model allowed them to extend the original phishing pages to increase game contents. Finally, the authors developed an anti-phishing educational game to evaluate the effectiveness of proposed frame hierarchy. The evaluation results showed that most of the lecturers and experts were satisfied with this proposed system.

Previous research has revealed that technology alone is insufficient to ensure critical IT security issues. So far, there has been little work on end user behaviour of performing security and preventing users from attacks which are imperative to cope up with MMFs such as phishing attacks \cite{arachchilage2016phishing} \cite{arachchilage2014security} \cite{arachchilage2013game} \cite{liang2009avoidance} \cite{liang2010understanding} \cite{workman2008security} \cite{aytes2004computer}. Many discussions have terminated with the conclusion of ``if we could only remove the user from the system, we would be able to make it secure" \cite{gorling2006myth}.  Where it is not possible to completely eliminate the user, for example in home use, the best possible approach for computer security is to educate the user in security prevention \cite{arachchilage2016phishing} \cite{arachchilage2014security} \cite{arachchilage2013game} \cite{kirlappos2012security} \cite{schneier2000semantic}. Previous research has revealed well designed user security education can be effective \cite{sheng2007anti} \cite{kumaraguru2008lessons} \cite{kumaraguru2007protecting} \cite{le2015renewed}. This could be web-based training materials, contextual training, and embedded training to improve users' ability to avoid phishing attacks. One objective of our research is to find effective ways to educate people to identify and prevent from MMFs  such as phishing websites. 

Therefore, this research focuses on investigating how one can better educate the people in order to protect themselves from phishing attacks. To address the problem, this research attempts to understand how people's ``self-efficacy" enhances their phishing threat avoidance behaviour. The research work reported in this paper then discusses how one can integrate people's ``self-efficacy'', into an anti-phishing educational game design in order to better educate themselves against phishing attacks. 

\section{Game Design Issues}

The aim of the proposed game design is to integrate people's ``self-efficacy'' into an anti-phishing educational game design in order to better educate themselves to thwart phishing attacks. Self-efficacy has a co-relation with individuals' knowledge \cite{hu2010self}. For example, uses are more confident to take relevant actions to thwart phishing attacks, when they are knowledgeable of phishing threats. McCormick \cite{mccormick1997conceptual} has revealed that one's knowledge can be influenced by learning procedural and conceptual knowledge associated with. Plant \cite{plant1994science} has argued that conceptual knowledge is close to the idea of ``know that" and procedural knowledge ``know how", in which, both the ideas are imperative to educate one to thwart phishing attacks. Furthermore, his research work described that such conceptual knowledge permits one to explain why, hence the difference of “know how” and “know why”. Additionally, McCormick \cite{mccormick1997conceptual} stated that the two ideas of conceptual and procedural knowledge are frequently treated individually, with their relationship being disregarded.

Therefore, in this research, the elements derived from a theoretical model \cite{arachchilage2014security} will be used to incorporate into the game design to thwart phishing attacks. The theoretical model \cite{arachchilage2014security} (shown in Fig. 2) examined whether conceptual knowledge or procedural knowledge effects on computer users' self-efficacy to thwart phishing attacks. Their findings revealed that the interaction effect of both procedural and conceptual knowledge, will positively influence on self-efficacy, which contributes to enhance computer users' phishing threat avoidance behaviour (through their motivation).

\begin{figure} 
\centering
\includegraphics [height=2.25 in,width=3.6 in]{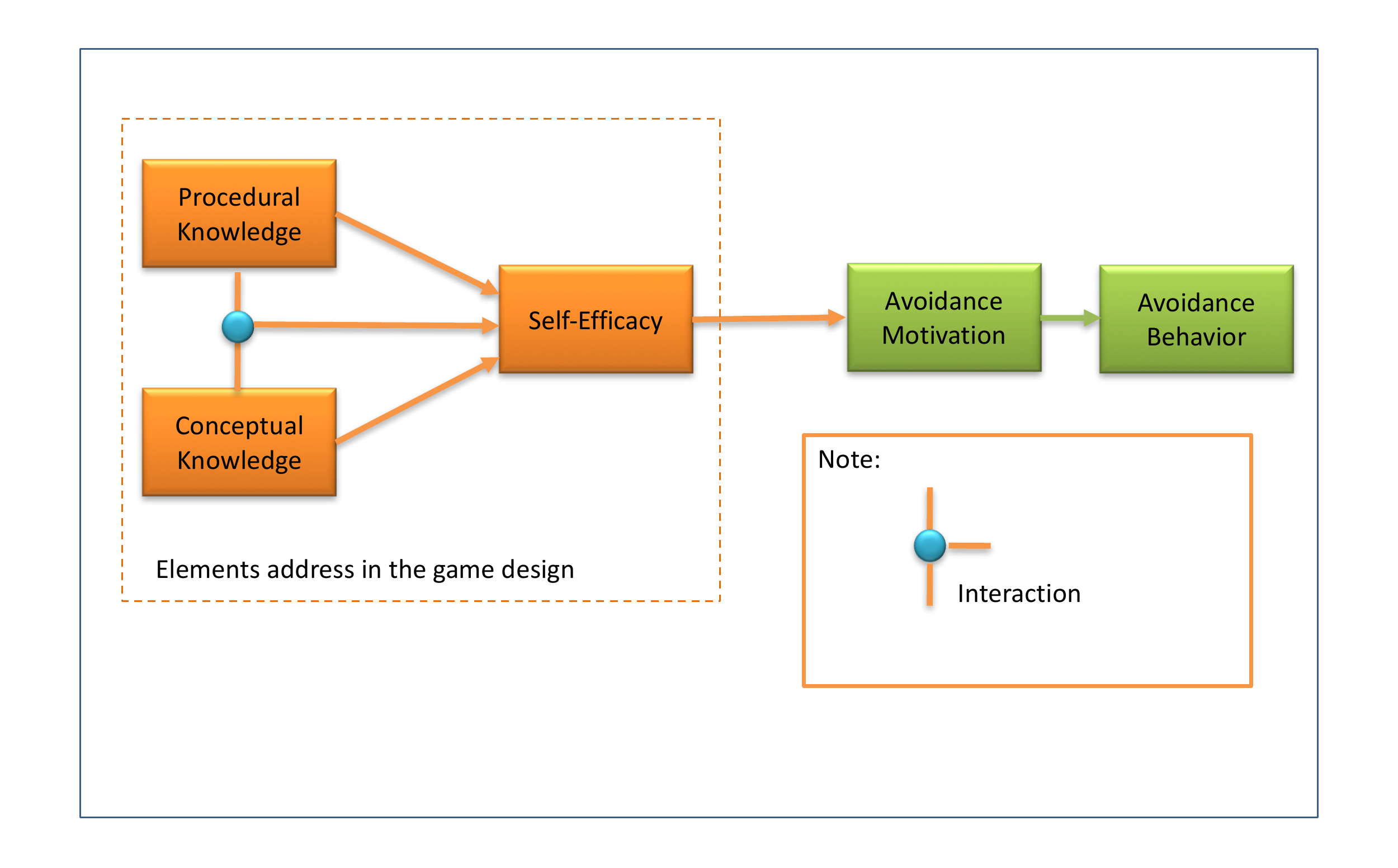}
\caption{ Elements address in the game design \cite{arachchilage2014security} \protect}
\vskip -6pt
\end{figure}  

\section{Integrating ``self-efficacy'' into a Game Design}

To explore the viability of using a game to thwart phishing attacks based on ``self-efficacy'', a prototype designed was proposed. We developed a story addressing ``self-efficacy'' (in this case, both procedural and conceptual knowledge as well as its interaction effect) of phishing URLs within a game design context. 

\subsubsection{Story}

The game story was created based on a scenario of a character of a small and big fish that both live in a big pond. The game player roll-plays the character of the small fish, who wants to eat worms in order to become a big fish. Worms were randomly generated in the game design. However, the small fish should be very careful of phishers, those who may try to trick him with fake worms. This represents phishing attacks by developing threat perception. The other character is the big fish, who is an experienced and grown up fish in the pond. The proposed mobile game design prototype contains two sections: teaching the concept of phishing URLs and phishing emails. The proposed game design teaches how one can differentiate legitimate URLs (Uniform Resource Locators) from fraudulent ones. 

\subsubsection{Game Design}

Each worm is associated with a URL, which appears as a dialog box. The small fish's job is to eat all the real worms, which are associated with legitimate URLs while prevent eating fake worms which are associated with fraudulent URLs before the time is up. If the phishing URL is correctly identified, then the game player is prompted to identify which part of the URL indicates phishing (shown in Fig. 3). This determines whether or not the game player has understood the conceptual knowledge of the phishing URL. At this point in time, the score of the game player will be increased to encourage him/her in order to proceed with the game. Nevertheless, if the phishing URL is incorrectly identified, then the game player will get real time feedback saying why their decision was wrong with an appropriate example, such as ``Legitimate websites usually do not have numbers at the beginning of their URLs". For example, http://187.52.91.111/.www.hsbc.co.uk”. Therefore, this attempts to teach the player about conceptual knowledge of phishing URLs within the game design context.

\begin{figure} 
\centering
\includegraphics [height=1.00 in,width=3.25 in]{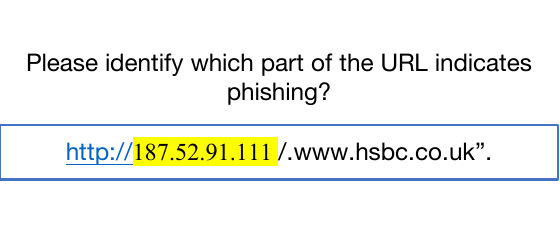}
\caption{ The game player is prompted to identify which part of the URL indicates phishing\protect}
\vskip -6pt
\end{figure}

Furthermore, one can argue that presenting the player with different types of URLs to identify if it is phishing or legitimate (i.e. procedural knowledge) then asking them to identify which part of the URL is phishing (i.e. conceptual knowledge), would positively impact on his/her self-efficacy to enhance the phishing threat avoidance.

The proposed game design is presented at different levels such as beginner, intermediate and advance. When the player comes across from the beginner to advanced level, complexity of the combination of URLs (i.e. procedural knowledge) is dramatically increased while considerably decreasing the time period to complete the game. Procedural knowledge of phishing prevention will be then addressed in the game design as the complexity of the URLs presented to the player through the design. The overall game design focuses on designing an innovative and gamified approach to integrate both conceptual and procedural knowledge effects on the game plyer's (i.e. computer users) self-efficacy to thwart phishing attacks.  

If the worm associated with the URL is suspicious or if it is difficult to identify/guess, the small fish (in this case the game player) can go to the big fish and request some help. The big fish will then provide some tips on how to recognize bad worms associated with fraudulent URLs. For example, ``a company name followed by a hyphen in a URL is generally a scam" or ``website addresses associated with numbers in the front are generally scams". This will again aid to develop the player's ``self-efficacy'' to combat against phishing attacks. Whenever the small fish asks some help from the big fish, the time left will be reduced by a certain amount (e.g. 100 seconds) as a payback for safeguarding measure.

\section{Conclusion and Future Work}

This research focused on designing an innovative and gamified approach to integrate both conceptual and procedural knowledge effects on the game player's (i.e. computer users) self-efficacy to combat against phishing attacks. The study asks how one can integrate ``self-efficacy" into an anti-phishing educational gaming tool that teach people to thwart phishing attacks? The game design teaches how one can differentiate legitimate URLs (Uniform Resource Locators) from fraudulent ones. The elements derived from a theoretical model \cite{arachchilage2014security} incorporated into the proposed game design to thwart phishing attacks. The theoretical model \cite{arachchilage2014security} revealed the interaction effects of both conceptual knowledge or procedural knowledge will positively effect on self-efficacy, which contributes to enhance computer users' phishing threat avoidance behaviour (through their motivation). Furthermore, as future research, we attempt to implement this game design and empirically investigate through users in order to understand how their knowledge (the interaction effect of both conceptual and procedural) will positively effect on self-efficacy, which eventually contributes to enhance their phishing threat avoidance behaviour.

The transformative nature of this proposal is in presenting the game based awareness design for individuals to protect themselves against phishing crimes. The proposed game designed that integrated the user's ``self-efficacy'' will enhance individual's phishing threat avoidance behaviour. This concept is based on the notion that not only can a computer game provides anti-phishing education, but also games potentially provide a better learning environment, because they motivate the user and keep attention by providing immediate feedback. This can also be considered as an appropriate way to reach individuals in society with the message of phishing threat awareness is vital, which makes a considerable contribution to enable the cyberspace a safe environment for everyone.

Eventually, the implemented game will enable the Australian government to protect their citizen being victims of phishing crimes. For example, the Australian government (for example, the office of the children's e-safety commissioner) can also make this game toolkit freely available to the Australian citizens in order to protect them from phishing crimes. Additionally, this proposed game design can be an enormous benefit to Australian schools to educate their children to protect themselves from phishing crimes. The proposed research work is also focused on integrating new knowledge and technology to provide law enforcement and security agencies with automatic devices and capabilities to improve prevention, detection and solution of crimes, and acts of terrorism.





%

\bibliographystyle{IEEEtran}
\bibliography{references}

\begin{thebibliography}{10}
\providecommand{\url}[1]{#1}
\csname url@samestyle\endcsname
\providecommand{\newblock}{\relax}
\providecommand{\bibinfo}[2]{#2}
\providecommand{\BIBentrySTDinterwordspacing}{\spaceskip=0pt\relax}
\providecommand{\BIBentryALTinterwordstretchfactor}{4}
\providecommand{\BIBentryALTinterwordspacing}{\spaceskip=\fontdimen2\font plus
\BIBentryALTinterwordstretchfactor\fontdimen3\font minus
  \fontdimen4\font\relax}
\providecommand{\BIBforeignlanguage}[2]{{%
\expandafter\ifx\csname l@#1\endcsname\relax
\typeout{** WARNING: IEEEtran.bst: No hyphenation pattern has been}%
\typeout{** loaded for the language `#1'. Using the pattern for}%
\typeout{** the default language instead.}%
\else
\language=\csname l@#1\endcsname
\fi
#2}}
\providecommand{\BIBdecl}{\relax}
\BIBdecl

\bibitem{Bitly2017}
\BIBentryALTinterwordspacing
Bitly, ``Bitly url shortener and link management platform,''
  \url{https://bitly.com}, October 2017, [Online: Accessed 15-June-2017].
  [Online]. Available: \url{https://bitly.com}
\BIBentrySTDinterwordspacing

\bibitem{Krawchenko2016}
K.~Krawchenko, ``The phishing email that hacked the account of john podesta,''
  \emph{CBS NEWS}, October 2016.

\bibitem{arachchilage2016phishing}
N.~A.~G. Arachchilage, S.~Love, and K.~Beznosov, ``Phishing threat avoidance
  behaviour: An empirical investigation,'' \emph{Computers in Human Behavior},
  vol.~60, pp. 185--197, 2016.

\bibitem{arachchilage2013game}
N.~A.~G. Arachchilage and S.~Love, ``A game design framework for avoiding
  phishing attacks,'' \emph{Computers in Human Behavior}, vol.~29, no.~3, pp.
  706--714, 2013.

\bibitem{gupta2017defending}
B.~Gupta, N.~A. Arachchilage, and K.~E. Psannis, ``Defending against phishing
  attacks: taxonomy of methods, current issues and future directions,''
  \emph{Telecommunication Systems}, pp. 1--21.

\bibitem{whitty2015mass}
M.~T. Whitty, ``Mass-marketing fraud: a growing concern,'' \emph{IEEE Security
  \& Privacy}, vol.~13, no.~4, pp. 84--87, 2015.

\bibitem{sheng2007anti}
S.~Sheng, B.~Magnien, P.~Kumaraguru, A.~Acquisti, L.~F. Cranor, J.~Hong, and
  E.~Nunge, ``Anti-phishing phil: the design and evaluation of a game that
  teaches people not to fall for phish,'' in \emph{Proceedings of the 3rd
  symposium on Usable privacy and security}.\hskip 1em plus 0.5em minus
  0.4em\relax ACM, 2007, pp. 88--99.

\bibitem{dhamija2006phishing}
R.~Dhamija, J.~D. Tygar, and M.~Hearst, ``Why phishing works,'' in
  \emph{Proceedings of the SIGCHI conference on Human Factors in computing
  systems}.\hskip 1em plus 0.5em minus 0.4em\relax ACM, 2006, pp. 581--590.

\bibitem{arachchilage2013taxonomy}
N.~A.~G. Arachchilage, C.~Namiluko, and A.~Martin, ``A taxonomy for securely
  sharing information among others in a trust domain,'' in \emph{Internet
  technology and secured transactions (ICITST), 2013 8th international
  conference for}.\hskip 1em plus 0.5em minus 0.4em\relax IEEE, 2013, pp.
  296--304.

\bibitem{arachchilage2014security}
N.~A.~G. Arachchilage and S.~Love, ``Security awareness of computer users: A
  phishing threat avoidance perspective,'' \emph{Computers in Human Behavior},
  vol.~38, pp. 304--312, 2014.

\bibitem{kirlappos2012security}
I.~Kirlappos and M.~A. Sasse, ``Security education against phishing: A modest
  proposal for a major rethink,'' \emph{IEEE Security \& Privacy}, vol.~10,
  no.~2, pp. 24--32, 2012.

\bibitem{kumaraguru2008lessons}
P.~Kumaraguru, S.~Sheng, A.~Acquisti, L.~F. Cranor, and J.~Hong, ``Lessons from
  a real world evaluation of anti-phishing training,'' in \emph{eCrime
  Researchers Summit, 2008}.\hskip 1em plus 0.5em minus 0.4em\relax IEEE, 2008,
  pp. 1--12.

\bibitem{walls2012using}
R.~Walls, ``Using computer games to teach social studies,'' Ph.D. dissertation,
  Uppsala University, Disciplinary Domain of Humanities and Social Sciences,
  Faculty of Educational Sciences, Department of Education.,
  http://www.diva-portal.org/smash/record.jsf?pid=diva2
  2012.

\bibitem{cone2007video}
B.~D. Cone, C.~E. Irvine, M.~F. Thompson, and T.~D. Nguyen, ``A video game for
  cyber security training and awareness,'' \emph{computers \& security},
  vol.~26, no.~1, pp. 63--72, 2007.

\bibitem{foreman2004video}
J.~Foreman, ``Video game studies and the emerging instructional revolution,''
  \emph{Innovate: Journal of Online Education}, vol.~1, no.~1, 2004.

\bibitem{arachchilage2012security}
G.~Arachchilage and N.~Asanka, ``Security awareness of computer users: A game
  based learning approach,'' Ph.D. dissertation, Brunel University, School of
  Information Systems, Computing and Mathematics, 2012.

\bibitem{zyda2005visual}
M.~Zyda, ``From visual simulation to virtual reality to games,''
  \emph{Computer}, vol.~38, no.~9, pp. 25--32, 2005.

\bibitem{michael2005serious}
D.~R. Michael and S.~L. Chen, \emph{Serious games: Games that educate, train,
  and inform}.\hskip 1em plus 0.5em minus 0.4em\relax Muska \&
  Lipman/Premier-Trade, 2005.

\bibitem{froschauer2010design}
J.~Froschauer, I.~Seidel, M.~G{\"a}rtner, H.~Berger, and D.~Merkl, ``Design and
  evaluation of a serious game for immersive cultural training,'' in
  \emph{Virtual Systems and Multimedia (VSMM), 2010 16th International
  Conference on}.\hskip 1em plus 0.5em minus 0.4em\relax IEEE, 2010, pp.
  253--260.

\bibitem{arachchilage2011design}
N.~A.~G. Arachchilage and M.~Cole, ``Design a mobile game for home computer
  users to prevent from ``phishing attacks'','' in \emph{Information Society
  (i-Society), 2011 International Conference on}.\hskip 1em plus 0.5em minus
  0.4em\relax IEEE, 2011, pp. 485--489.

\bibitem{bellotti2009enhancing}
F.~Bellotti, R.~Berta, A.~D. Gloria, and L.~Primavera, ``Enhancing the
  educational value of video games,'' \emph{Computers in Entertainment (CIE)},
  vol.~7, no.~2, p.~23, 2009.

\bibitem{gustafsson2009evaluation}
A.~Gustafsson, C.~Katzeff, and M.~Bang, ``Evaluation of a pervasive game for
  domestic energy engagement among teenagers,'' \emph{Computers in
  Entertainment (CIE)}, vol.~7, no.~4, p.~54, 2009.

\bibitem{eagle2008wu}
M.~Eagle and T.~Barnes, ``Wu's castle: teaching arrays and loops in a game,''
  in \emph{ACM SIGCSE Bulletin}, vol.~40, no.~3.\hskip 1em plus 0.5em minus
  0.4em\relax ACM, 2008, pp. 245--249.

\bibitem{gee2003video}
J.~P. Gee, ``What video games have to teach us about learning and literacy,''
  \emph{Computers in Entertainment (CIE)}, vol.~1, no.~1, pp. 20--20, 2003.

\bibitem{fotouhi2009mobo}
F.~Fotouhi-Ghazvini, R.~Earnshaw, D.~Robison, and P.~Excell, ``The mobo city: A
  mobile game package for technical language learning.'' \emph{International
  Journal of Interactive Mobile Technologies}, vol.~3, no.~2, 2009.

\bibitem{gunter2008taking}
G.~A. Gunter, R.~F. Kenny, and E.~H. Vick, ``Taking educational games
  seriously: using the retain model to design endogenous fantasy into
  standalone educational games,'' \emph{Educational Technology Research and
  Development}, vol.~56, no. 5-6, pp. 511--537, 2008.

\bibitem{malone1981toward}
T.~W. Malone, ``Toward a theory of intrinsically motivating instruction,''
  \emph{Cognitive science}, vol.~5, no.~4, pp. 333--369, 1981.

\bibitem{oblinger2004next}
D.~Oblinger, ``The next generation of educational engagement,'' \emph{Journal
  of interactive media in education}, vol. 2004, no.~1, 2004.

\bibitem{squire2003video}
K.~Squire, ``Video games in education,'' in \emph{International journal of
  intelligent simulations and gaming}.\hskip 1em plus 0.5em minus 0.4em\relax
  Citeseer, 2003.

\bibitem{de2006can}
S.~De~Freitas and M.~Oliver, ``How can exploratory learning with games and
  simulations within the curriculum be most effectively evaluated?''
  \emph{Computers \& education}, vol.~46, no.~3, pp. 249--264, 2006.

\bibitem{prensky2001digital}
M.~Prensky, ``Digital game-based learning,'' 2001.

\bibitem{moreno2008educational}
P.~Moreno-Ger, D.~Burgos, I.~Mart{\'\i}nez-Ortiz, J.~L. Sierra, and
  B.~Fern{\'a}ndez-Manj{\'o}n, ``Educational game design for online
  education,'' \emph{Computers in Human Behavior}, vol.~24, no.~6, pp.
  2530--2540, 2008.

\bibitem{andrews2003concept}
G.~Andrews, E.~Woodruff, K.~A. MacKinnon, and S.~Yoon, ``Concept development
  for kindergarten children through a health simulation,'' \emph{Journal of
  Computer Assisted Learning}, vol.~19, no.~2, pp. 209--219, 2003.

\bibitem{roschelle2002walk}
J.~Roschelle and R.~Pea, ``A walk on the wild side: How wireless handhelds may
  change computer-supported collaborative learning,'' \emph{International
  Journal of Cognition and Technology}, vol.~1, no.~1, pp. 145--168, 2002.

\bibitem{allen2006social}
M.~Allen, ``Social engineering: A means to violate a computer system,''
  \emph{SANS Institute, InfoSec Reading Room}, 2006.

\bibitem{schechter2007emperor}
S.~E. Schechter, R.~Dhamija, A.~Ozment, and I.~Fischer, ``The emperor's new
  security indicators,'' in \emph{Security and Privacy, 2007. SP'07. IEEE
  Symposium on}.\hskip 1em plus 0.5em minus 0.4em\relax IEEE, 2007, pp. 51--65.

\bibitem{timko2008social}
D.~Timko, ``The social engineering threat,'' \emph{Information Systems Security
  Association Journal}, 2008.

\bibitem{herley2009so}
C.~Herley, ``So long, and no thanks for the externalities: the rational
  rejection of security advice by users,'' in \emph{Proceedings of the 2009
  workshop on New security paradigms workshop}.\hskip 1em plus 0.5em minus
  0.4em\relax ACM, 2009, pp. 133--144.

\bibitem{baslyman2016smells}
M.~Baslyman and S.~Chiasson, ``" smells phishy?": An educational game about
  online phishing scams,'' in \emph{Electronic Crime Research (eCrime), 2016
  APWG Symposium on}.\hskip 1em plus 0.5em minus 0.4em\relax IEEE, 2016, pp.
  1--11.

\bibitem{kumaraguru2007protecting}
P.~Kumaraguru, Y.~Rhee, A.~Acquisti, L.~F. Cranor, J.~Hong, and E.~Nunge,
  ``Protecting people from phishing: the design and evaluation of an embedded
  training email system,'' in \emph{Proceedings of the SIGCHI conference on
  Human factors in computing systems}.\hskip 1em plus 0.5em minus 0.4em\relax
  ACM, 2007, pp. 905--914.

\bibitem{robila2006don}
S.~A. Robila and J.~W. Ragucci, ``Don't be a phish: steps in user education,''
  in \emph{ACM SIGCSE Bulletin}, vol.~38, no.~3.\hskip 1em plus 0.5em minus
  0.4em\relax ACM, 2006, pp. 237--241.

\bibitem{jagatic2007social}
T.~N. Jagatic, N.~A. Johnson, M.~Jakobsson, and F.~Menczer, ``Social
  phishing,'' \emph{Communications of the ACM}, vol.~50, no.~10, pp. 94--100,
  2007.

\bibitem{tseng2011automatic}
S.-S. Tseng, K.-Y. Chen, T.-J. Lee, and J.-F. Weng, ``Automatic content
  generation for anti-phishing education game,'' in \emph{Electrical and
  Control Engineering (ICECE), 2011 International Conference on}.\hskip 1em
  plus 0.5em minus 0.4em\relax IEEE, 2011, pp. 6390--6394.

\bibitem{liang2009avoidance}
H.~Liang and Y.~Xue, ``Avoidance of information technology threats: a
  theoretical perspective,'' \emph{MIS quarterly}, pp. 71--90, 2009.

\bibitem{liang2010understanding}
------, ``Understanding security behaviors in personal computer usage: A threat
  avoidance perspective,'' \emph{Journal of the Association for Information
  Systems}, vol.~11, no.~7, p. 394, 2010.

\bibitem{workman2008security}
M.~Workman, W.~H. Bommer, and D.~Straub, ``Security lapses and the omission of
  information security measures: A threat control model and empirical test,''
  \emph{Computers in human behavior}, vol.~24, no.~6, pp. 2799--2816, 2008.

\bibitem{aytes2004computer}
K.~Aytes and T.~Connolly, ``Computer security and risky computing practices: A
  rational choice perspective,'' \emph{Journal of Organizational and End User
  Computing (JOEUC)}, vol.~16, no.~3, pp. 22--40, 2004.

\bibitem{gorling2006myth}
S.~Gorling, ``The myth of user education,'' 2006.

\bibitem{schneier2000semantic}
B.~Schneier, ``Semantic attacks: The third wave of network attacks,''
  \emph{Crypto-Gram Newsletter}, vol.~14, 2000.

\bibitem{le2015renewed}
A.~Le~Compte, D.~Elizondo, and T.~Watson, ``A renewed approach to serious games
  for cyber security,'' in \emph{Cyber conflict: Architectures in cyberspace
  (CyCon), 2015 7th international conference on}.\hskip 1em plus 0.5em minus
  0.4em\relax IEEE, 2015, pp. 203--216.

\bibitem{hu2010self}
W.~Hu, ``Self-efficacy and individual knowledge sharing,'' in \emph{Information
  Management, Innovation Management and Industrial Engineering (ICIII), 2010
  International Conference on}, vol.~2.\hskip 1em plus 0.5em minus 0.4em\relax
  IEEE, 2010, pp. 401--404.

\bibitem{mccormick1997conceptual}
R.~McCormick, ``Conceptual and procedural knowledge,'' in \emph{Shaping
  Concepts of Technology}.\hskip 1em plus 0.5em minus 0.4em\relax Springer,
  1997, pp. 141--159.

\bibitem{plant1994science}
M.~Plant, ``How is science useful to technology,'' \emph{Design and Technology
  in the Secondary Curriculum: A Book of Readings, The Open University, Milton
  Keynes}, pp. 96--108, 1994.

\end{thebibliography}


\begin{thebibliography}{1}

\bibitem{IEEEhowto:kopka}
H.~Kopka and P.~W. Daly, \emph{A Guide to \LaTeX}, 3rd~ed.\hskip 1em plus
  0.5em minus 0.4em\relax Harlow, England: Addison-Wesley, 1999.

\end{thebibliography}

\end{document}